\documentclass[prd,nofootinbib,preprint,superscriptaddress,twocolumn,10pt]{revtex4}
\pdfoutput=1
\usepackage{amsmath}
\usepackage{float}
\usepackage{amsfonts}
\usepackage{amssymb}
\usepackage{mathrsfs}
\usepackage{cancel}
\usepackage{accents}
\usepackage{mciteplus,slashed}
\usepackage{amssymb,cancel,amsmath,relsize}
\usepackage{mathrsfs} 
\usepackage{dcolumn}
\usepackage{bm}
\usepackage[caption=false]{subfig} 
\usepackage{appendix}
\usepackage{physics}
\usepackage{feynmp-auto}
\usepackage[T1]{fontenc}	
\usepackage{csvsimple}
\usepackage{hyperref}
\usepackage[capitalise]{cleveref}
\usepackage{booktabs}
\usepackage{graphicx}
\usepackage{mathrsfs}
\usepackage[utf8]{inputenc}
\usepackage[T1]{fontenc}
\usepackage[dvipsnames]{xcolor}

\hypersetup{
    colorlinks,
    linkcolor={red!50!black},
    citecolor={blue!50!black},
    urlcolor={blue!80!black}
}
\usepackage[normalem]{ulem}
\usepackage{cleveref}

\renewcommand{\section}[1]{{\noindent \bf{#1.}---}}

\newcommand{\Ztwo}{\ensuremath{\mathcal{Z}_2}}
\newcommand{\ZDM}{\ensuremath{\mathcal{Z}_2^{\rm DM}}}
\newcommand{\ZDW}{\ensuremath{\mathcal{Z}_2^{\rm DW}}}

\newcommand{\bmt}{\begin{pmatrix}}
\newcommand{\emt}{\end{pmatrix}}
\newcommand{\ba}{\begin{array}{c}}
\newcommand{\ea}{\end{array}}
\newcommand{\be}{\begin{equation}}
\newcommand{\ee}{\end{equation}}
\newcommand{\bea}{\begin{eqnarray}}
\newcommand{\eea}{\end{eqnarray}}

\newcommand{\bi}{\begin{itemize}}
\newcommand{\ei}{\end{itemize}}

\newcommand{\baz}{\begin{array}{cc}}
\newcommand{\besub}{\begin{subequations}}
\newcommand{\eesub}{\end{subequations}}

\newcommand{\mathsym}[1]{{}}

\newcommand{\bt}{\begin{tabular}}
\newcommand{\et}{\end{tabular}}

\newcommand{\benu}{\begin{enumerate}}
\newcommand{\eenu}{\end{enumerate}}

\def\l{\lambda}

\def\q2 {q^2}

\def\bt{\begin{table}}
\def\et{\end{table}}

\newcommand{\bav}{\begin{array}{cccc}}

\graphicspath{{Figures/}}

\begin{document}

\title{ Twin-peaked gravitational wave signals from a $\mathcal{Z}_2$ phase transition}


\author{Rishav Roshan}
\affiliation{School of Physics and Astronomy, University of Southampton, Southampton SO17 1BJ, United Kingdom}
\author{Indrajit Saha}
\affiliation{Department of Physics, Indian Institute of Technology Guwahati, Assam 781039, India}

\begin{abstract}

We compute the gravitational wave spectrum from a phase transition associated with the spontaneous breaking of a $\ZDW$ symmetry. If the transition is second-order, the only source of gravitational waves is the annihilation of domain walls (biased by quantum gravity) formed after this breaking. However, if the transition is first-order, this yields a twin-peaked signal from both the transition itself and the biased domain wall annihilation. Both scenarios originate when a scalar singlet odd under the $\ZDW$ obtains a non-zero vacuum expectation value. An additional $\ZDM$ odd scalar doublet strengthens the transition by keeping the singlet scalar in thermal equilibrium with the Standard Model plasma at early times. Additionally, the same scalar doublet produces fermionic dark matter via freeze-in, matching observed dark matter relic density.

\end{abstract}

\maketitle
\tableofcontents
\section{Introduction}
Gravity being the weakest fundamental force allows gravitons to decouple from the thermal plasma at the Planck epoch, making gravitational waves (GWs) a unique messenger capable of carrying unperturbed information from the very beginning of the Universe. Moreover, the recent discovery of stochastic gravitational wave background (SGWB), as reported by the North American Nanohertz Observatory for Gravitational Waves (NANOGrav)~\cite{NANOGrav:2023gor, NANOGrav:2023hvm} now provides a concrete opportunity to exploit this feature, motivating exploration of cosmological events from the early Universe that could generate such a background and potentially reveal physics beyond the Standard Model (SM).

A SGWB may originate from a variety of early cosmological events, such as first-order phase transitions (FOPT)~\cite{Winicour1973, Hogan:1986qda, Athron:2023xlk, Caprini:2010xv, NANOGrav:2021flc, Xue:2021gyq, DiBari:2021dri, Madge:2023cak}, cosmic strings~\cite{Siemens:2006yp, Ellis:2020ena, King:2020hyd, Buchmuller:2020lbh, Blasi:2020mfx, Bian:2022tju, Fu:2023nrn, Mandal:2026wbz}, or domain walls (DWs) annihilation~\cite{Ferreira:2022zzo, An:2023idh, Dunsky:2021tih, Sakharov:2021dim,Bai:2023cqj,Roshan:2026yon}, primordial black holes \cite{Domenech:2024kmh, Flores:2024eyy, del-Corral:2025fca, Gross:2024wkl, Borah:2022vsu,Barman:2022pdo,Borah:2023iqo,Barman:2024slw,Borah:2024lml}, inflation \cite{Grishchuk:1974ny,Starobinsky:1979ty,Guzzetti:2016mkm, Barman:2024ujh} etc. While most of these studies have focused on GW production from individual sources, understanding the complete GW spectrum observable today requires considering them within a comprehensive theoretical framework. With this in mind, we investigate the GW spectrum generated by the annihilation of DWs that form after the Universe undergoes a phase transition. 

DWs are a two dimensional sheet-like topological defects that arise from the spontaneous breaking of a discrete symmetry \cite{Vilenkin:1984ib,Kolb:1990vq,Preskill:1992ck,1994csot.book.....V}. The study of DWs has traditionally focused on second-order phase transitions (SOPTs), where they appear after a smooth phase transition, an idea originally introduced by T. D. Lee~\cite{Lee:1974jb}. However, if the discrete symmetry is spontaneously broken after the Universe undergoes a FOPT, the formation and evolution of DWs exhibit distinct characteristics that need a separate investigation as also discussed in \cite{Wei:2022poh,Borboruah:2022eex,Fornal:2024avx}. Regardless of the formation mechanism, DWs' energy density dilutes slowly ($\rho_{\rm DW}\propto t^{-1}$) compared to radiation and matter, causing them to rapidly dominate the energy budget of the Universe. The slower dilution can lead to catastrophic cosmological consequences. In order to avoid such catastrophe, the DWs can be made unstable by explicitly breaking the discrete symmetries \cite{Gelmini:1988sf} and introducing biased conditions at the time of formation\footnote{However, this bias can also be generated through radiative corrections~\cite{Zhang:2023nrs, Zeng:2025zjp, Borah:2025bfa}.}. This allows the DW network to annihilate prior to dominating the Universe's energy budget, efficiently producing GWs in the process~\cite{Saikawa:2017hiv,Roshan:2024qnv}. While DW annihilation remains the sole source of the SGWB in SOPT scenarios, the situation becomes richer after a FOPT, where GWs arise from two distinct mechanisms: (i) the annihilation of DWs and (ii) the dynamics of the phase transition itself.

Building on our previous studies \cite{King:2023ayw,King:2023ztb,Borah:2024kfn,Gouttenoire:2025ofv}, in this work we extend the model by introducing an additional $SU(2)_L$ scalar doublet and a fermionic dark matter (DM) candidate. Since the scalar is an $SU(2)_L$ gauge doublet, its gauge interactions with the Standard Model particles keep it in thermal equilibrium with the plasma at high temperatures. Finally, the scalar doublet decay also produces the DM through freeze-in mechanism \cite{Hall:2009bx,Bernal:2017kxu, Biswas:2016bfo, Datta:2021elq,  Bhattacharya:2021jli,Elahi:2014fsa,Biswas:2019iqm,Barman:2020plp,Barman:2021tgt}. Following, \cite{King:2023ayw,King:2023ztb,Borah:2024kfn, Gouttenoire:2025ofv}, we also consider a scalar, singlet under the SM gauge symmetry. Both the scalars are considered to be odd under the discrete global $\ZDW$ symmetry, while the newly introduced scalar doublet additionally carries a nontrivial charge under $\ZDM$, which also stabilizes the DM odd under $\ZDM$. Note that, through its coupling to the scalar singlet, scalar doublet also maintains the singlet in thermal equilibrium with the SM bath, ensuring that both sectors share the same temperature. While the singlet scalar is responsible for breaking the $\ZDW$ symmetry and creating the DW, its interactions with the scalar doublet play a crucial role in determining the nature of the phase transition.

We follow the similar approach as discussed in \cite{King:2023ayw,King:2023ztb,Borah:2024kfn, Gouttenoire:2025ofv} and rely on the fact that  exact global symmetries should be explicitly broken \cite{Banks:1988yz,Banks:2010zn,Harlow:2018tng} in theories of quantum gravity (QG). These small but non-zero symmetry-breaking makes both DM and DWs unstable, allowing them to decay and produce observable signatures: indirect detection signals from DM decay and GW from DW annihilation. Our analysis proceeds in two stages. First, we consider the parameter space where the phase transition is second order, such that DW annihilation constitutes the only GW source. Second, we study GW generation when DWs form following a FOPT. This latter scenario generates a distinctive twin-peaked GW spectrum that can be simultaneously probed by multiple GW experiments.

The paper is organized as follows. In Section \ref{model} we introduce our simplified models involving two newly introduced scalars and a fermionic DM. In Section \ref{PT} we discuss the dynamics of phase transition and DW formation. In Section \ref{QG}, we discuss how QG effects can destabilise the DW networks followed by an analysis of GW generation in Section~\ref{GW}. Section~\ref{DMPH} addresses DM production and detection prospects. Finally, we conclude in Section~\ref{conclusion}.

\vspace{0.5 em} 

\section{ Framework}
\label{model}
\noindent In this work, we consider a setup where the SM is extended by a singlet scalar field $S$ and an additional $SU(2)_L$ scalar doublet $\eta$. While both the scalars are non-trivially charged under a discrete symmetry $\ZDW$, $\eta$ also carries a $-1$ unit of charge under $\ZDM$. This generates the following scalar potential at tree level
\begin{eqnarray}
    V_0&=& \mu ^2  H^\dagger H+ \lambda (H^\dagger H)^2 + \mu_\eta^2 \eta^{\dagger} \eta+ \l_\eta (\eta^{\dagger} \eta)^2 \\ && + \l_1 (H^{\dagger} H) (\eta^{\dagger} \eta)
 + \l_2 (H^{\dagger} \eta) (\eta^{\dagger} H)
 \nonumber \\ &&+ \frac{\l_3}{2} \Big[(H^{\dagger} \eta)^2 + 
 (\eta^{\dagger} H)^2 \Big] +\lambda _{hs}S^2 H^\dagger H  \nonumber \\ && 
    +\lambda _{\eta s}S^2 \eta^\dagger \eta  +\frac{\lambda_s}{4} (S^2-v_s^2)^2   \ ,
    \label{pot_tree}
\end{eqnarray}
where $H$ is the SM Higgs doublet field. Further we introduce a vector like lepton $\chi$, a singlet under the SM gauge symmetry but protected by an approximate $\ZDM$. As a result, one can write the following Lagrangian,
 \begin{equation}
     \mathcal{L}=y_\chi\bar{\ell}\tilde{\eta}\chi+m_\chi\bar{\chi}\chi+h.c.
 \end{equation}
 The scalar potential should be bounded from below to make the electroweak vacuum stable, which poses constraints on the scalar couplings ~\cite{Kannike:2012pe}. Once $\ZDW$ is spontaneously broken, the mass of $\eta$ can be expressed as $m_{\eta}^2=\mu_\eta^2+\lambda _{\eta s}v_s^2$. Additionally, a small quartic coupling $\lambda_{hs}$ between $S$ and the SM Higgs, together with the large hierarchy $v_{s}\gg v_h$, ensures that the mixing angle $\theta_{sh} \sim \lambda_{hs} v_h v_{s}/(m_{s}^2 - m^2_h)$ remains sufficiently small. This also prevent large quadratic correction to the Higgs mass, avoids observable deviations in Higgs phenomenology and allows us to neglect $S$-Higgs mixing in our analysis.

 Moreover, the two $\mathcal{Z}_2$-symmetries are also broken by QG effects by higher-dimensional operators. While the $\ZDW$ is broken by the operators of the form 
    \begin{eqnarray}
        \Delta V  &=& \frac{1}{\Lambda_{\rm QG}} (\alpha_{1}  S^5 + \alpha_{2h}  S^3 H^2 + \alpha_{3h}  S H^4 ) \nonumber \\ 
        && + \frac{1}{\Lambda_{\rm QG}} ( \alpha_{2\eta}  S^3 \eta^2 + \alpha_{3\eta}  S \eta^4 ) , 
        \label{eq:delV}
    \end{eqnarray}

\noindent both $\ZDW$ and $\ZDM$ symmetries are broken explicitly by,  
\begin{eqnarray}
        \mathcal{L}_{\text{break}}  &=& \sum_{i=1,2,3} \frac{ \beta_i}{\Lambda_{\rm QG}}  S\overline{\ell_{L_i}}\widetilde{H}\chi +h.c. \; ,
        \label{eq:Z2both}
    \end{eqnarray}   
where $\widetilde{H}:=i\sigma_2 H^*$.
 We also assume that all the global symmetries are broken by a common scale \cite{King:2023ayw,King:2023ztb,Borah:2024kfn, Gouttenoire:2025ofv}. This means that we simply take all the dimensionless coefficients $\alpha_i$ and $\beta_i$ \footnote{We also consider that the DM interacts with different generations of neutrinos with the identical ${\cal O}(1)$  strength, i.e. $\beta_1=\beta_2=\beta_3=1$.}
in the above equations to be of the same order and we can make all of them to be $\mathcal{O}(1)$ by redefining $\Lambda_{\rm QG}$.

It is worth noting that the cutoff scale $\Lambda_{\rm QG}$ associated with higher-dimensional operators that break the global symmetry is typically taken to be the reduced Planck mass $M_{\rm Pl}$ \cite{Addazi:2021xuf}. This choice is natural when symmetry breaking arises from perturbative gravitational effects. However, non-perturbative effects can substantially suppress the symmetry-breaking effects, yielding an effective cutoff scale that exceeds $M_{\rm Pl}$ by many orders of magnitude:
\begin{equation}
\Lambda_{\rm QG} = M_{\rm Pl} e^{\mathcal{S}} \gg M_{\rm Pl},
\end{equation}
where $\mathcal{S}$ denotes the instanton action.


\section{Dynamics of Phase Transition and domain wall formation}
\label{PT}
Next, we discuss the dynamics of phase transition in this model. To investigate the high-temperature behaviour of the scalar potential, we first evaluate the complete effective potential, incorporating both the tree-level potential and loop corrections at finite temperature. The effective scalar potential can be expressed as 

\begin{eqnarray}
    V_{\rm eff}&=&V_0+V_{\rm CW}+V_{\rm CT}+V_{\rm Th}+V_{\rm daisy},
    \label{Veff}
\end{eqnarray}

\noindent where $V_0,V_{\rm CW}$ and $V_{\rm CT}$ denotes the tree-level potential given in \eqref{pot_tree}, Coleman-Weinberg corrections and counter terms respectively and, $V_{\rm Th}$ and $V_{\rm daisy}$ denotes the leading thermal corrections to scalar potential \footnote{In analyzing the phase transition, we neglect the SM Higgs contribution to the scalar potential. This simplification is justified by the suppressed portal coupling $\lambda_{hs}$, discussed in Section \ref{model}.}  The Coleman-Weinberg (CW) potential \cite{Coleman:1973jx} can then be expressed as,

\begin{equation}
V_{\rm CW}=\frac{1}{64 \pi^2}\sum_{i}(-1)^{2s_i}n_i M_i^4(S)\left \{\log{ \left ( \frac{M_i^2(S)}{\mu^2} \right )}-C_i\right \},
\end{equation}
where the sum $i$ runs over the contributions from all particles in the theory, $s_i$ and $n_i$ are the spin and the number of degrees of freedom, respectively, with $n_{S}=1$ and $n_{\eta}=4$. $\mu$ denotes the renormalization scale and we fix it at $v_{ s}$. In this work, we adopt the $\overline{\rm MS}$ scheme with  $C_{i}=3/2$. Below we provide the field dependent masses for both the scalars
\begin{eqnarray}
    m_\eta^2(S) & =\mu_\eta^2 + \lambda_{\eta s}S^2, ~
    m_S^2(S) & = \lambda_s (3S^2-v_s^2)
    \end{eqnarray}

As pointed out in \cite{Coleman:1973jx}, the inclusion of the loop corrections shifts the minima of the scalar potential at the tree-level. In such a case, the counter terms are introduced to keep the minima intact. The counter term potential can be written as,
\begin{eqnarray}
    V_{\rm CT}&=& \frac{\delta\mu_s^2}{2} S^2+\frac{\delta \lambda_s}{4} S^4.
    \label{CT_Pot}
\end{eqnarray}
The above counter terms are determined by solving the following sets of equation,

\begin{eqnarray}
    \bigg(\frac{\partial V_{\rm CT}}{\partial S}+\frac{\partial V_{\rm CW}}{\partial S}\bigg)\bigg{|}_{v_s}=0, \nonumber \\
    \bigg(\frac{\partial^2 V_{\rm CT}}{\partial S^2}+\frac{\partial^2 V_{\rm CW}}{\partial S^2}\bigg)\bigg{|}_{v_s}=0.
    \label{CT}
\end{eqnarray}
Solving \eqref{CT}, we get,
\begin{eqnarray}
    \delta \mu_s^2&=&\frac{1}{2}\frac{\partial^2 V_{\rm CW}}{\partial S^2}-\frac{3}{2 v_s}\frac{\partial V_{\rm CW}}{\partial S}\nonumber\\
     \delta \lambda_s&=&\frac{1}{2 v_s^3}\frac{\partial V_{\rm CW}}{\partial S}-\frac{1}{2 v_s^2}\frac{\partial^2 V_{\rm CW}}{\partial S^2}
\end{eqnarray}
Now, since $S$ is in thermal equilibrium with the SM bath at high temperatures, the thermal contributions to the effective potential \cite{Dolan:1973qd,Quiros:1999jp} can be expressed as
\begin{equation}
V_{\rm Th}=\sum_i\frac{n_iT^4}{2\pi^2}J_B \left(\frac{M_i^2(S)}{T^2}\right)
\end{equation}
where\\
$$ J_B(x)=\int_{0}^{\infty}dy\, y^2 \log[1-e^{-\sqrt{y^2+x^2}}],$$
and $T$ is the temperature of the thermal plasma. Besides this, the Daisy corrections \cite{Fendley:1987ef,Parwani:1991gq,Arnold:1992rz} are also incorporated in thermal contribution to improve perturbative expansion considering Arnold-Espinosa method \cite{Arnold:1992rz} during FOPT such that $V_{\rm thermal}(S,T)=V_{\rm Th}(S,T) + V_{\rm daisy}(S,T)$. The Daisy contribution can be written as
\begin{equation}
V_{\rm daisy}(S,T)=-\frac{T}{2\pi^2}\sum_i n_i[M_i^3(S,T)-M_i^3(S)]
\end{equation}
The thermal corrections to the masses of the different particles involved can be expressed as~\cite{Wang:2019pet,Zhang:2021alu,Zhou:2022mlz} 

\begin{align}
    \Pi_\eta & = \left(\frac{g^2}{8} + \frac{(g^2+g'^2)}{16} + \frac{\lambda_\eta}{2} + \frac{\lambda_1+\lambda_2}{12} +\frac{\lambda_{\eta s}}{12} \right)T^2 \\
    \Pi_S & = \left(\frac{\lambda_s}{4}+\frac{\lambda_{\eta s}}{3} \right)T^2
\end{align}
 As the Universe expanded and cooled, the $\ZDW$ symmetry was spontaneously broken during a phase transition. The thermal evolution of the scalar field $S$ is governed by the finite-temperature effective potential $V_{\text{eff}}(S,T)$, which determines the preferred vacuum state at a given temperature. The shape of this potential evolves with temperature, and the transition from the symmetric phase $(\langle S \rangle = 0)$ to the broken phase $(\langle S \rangle \neq 0)$ corresponds to a cosmological phase transition.

%


\subsection{Second order phase transition and domain wall formation}
\begin{figure}[htbp!]
    \centering
    \includegraphics[width=\linewidth]{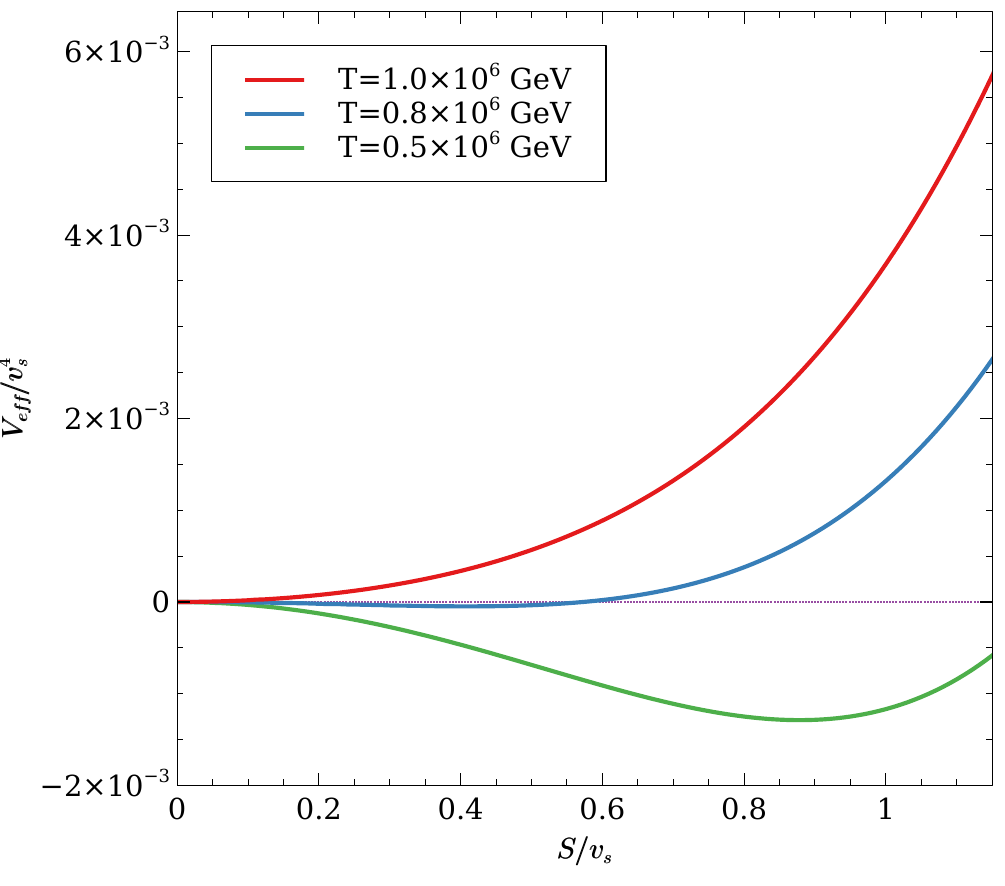}
    \caption{Temperature dependence of the effective potential $V_{\rm eff}(S,T)$ for SOPT. The potential profiles are shown for $\lambda_s = 0.01$, $\lambda_{\eta s} = 0.2$, and $v_s = 10^6~\mathrm{GeV}$.}
    \label{fig:sopt}
\end{figure}

In principle, the phase transition can be either second order or first order. For a SOPT, the Universe smoothly evolves from the symmetric phase to the broken phase as the temperature decreases. In this case, the minimum of the effective potential continuously shifts from $\langle S \rangle = 0$ to $\langle S \rangle \neq 0$. This behaviour is illustrated in the potential plots shown in Fig.~\ref{fig:sopt}, for the parameter choice $\lambda_s = 0.01$, $\lambda_{\eta s} = 0.2$, $\mu_\eta=10^{-2}v_s$ and $v_s = 10^6~\text{GeV}$.

Although the SOPT occurs continuously, causality prevents different regions of the Universe from choosing the same vacuum state. At the time of the transition, the Universe is divided into causally disconnected regions with size of order the correlation length $\xi (T)=1/m_S(T)$, where $m_S(T)$ is thermal mass. In each region, the scalar field independently settles into either $+v_s$ or $-v_s$ with roughly equal probability. When neighboring regions choose different vacua, the scalar field must interpolate between them. This interpolation cannot occur abruptly, instead, the field varies smoothly across a finite thickness region, forming a DW. Thus, DWs appear at the boundaries separating regions with different vacuum choices.

\subsection{First order phase transition and domain wall formation}
\label{DW_FOPT}

\begin{figure}[htbp!]
    \centering
    \includegraphics[width=\linewidth]{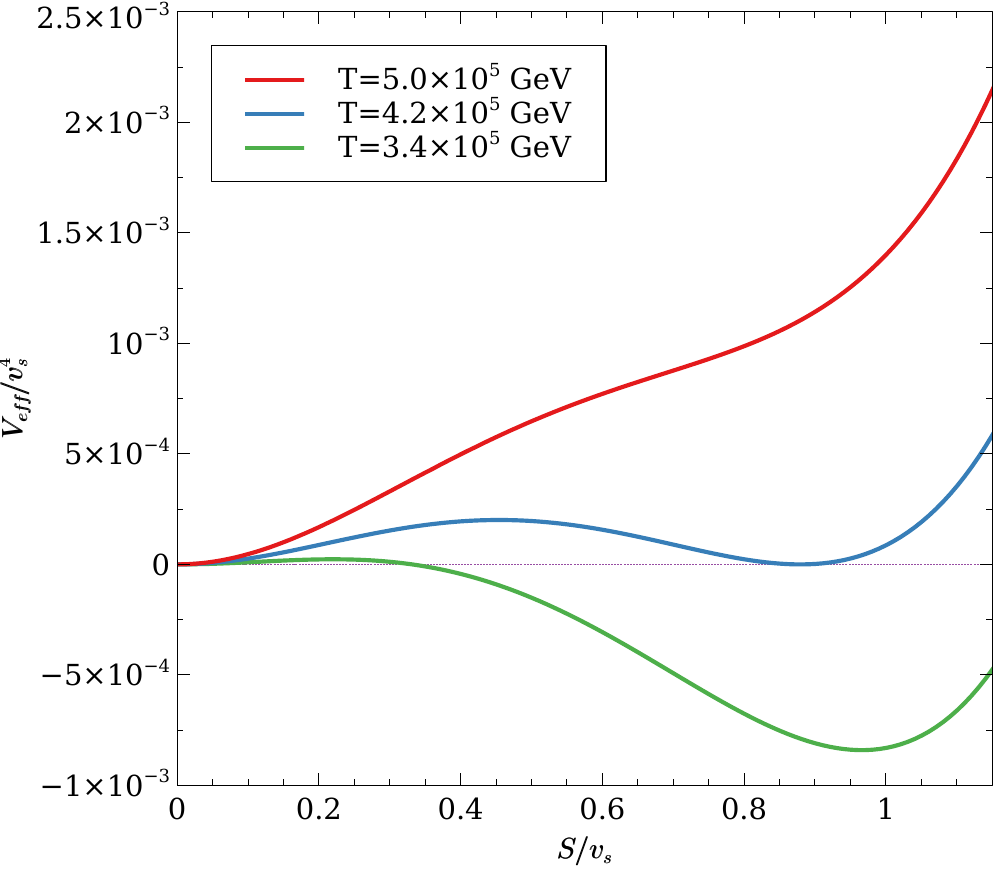}
    \caption{Temperature evolution of the effective scalar potential showing a FOPT. The potential profiles are shown for $\lambda_s = 0.01$, $\lambda_{\eta s} = 1.0$, and $v_s = 10^6~\mathrm{GeV}$.}
    \label{fig:fopt}
\end{figure}

For a FOPT, the effective potential develops a second, local minimum separated by a potential barrier from the symmetric phase. At the critical temperature $T_c$, the two minima become degenerate. The transition then proceeds through the nucleation of bubbles of the broken phase within the symmetric background. As a result, the order parameter (vacuum expectation value) changes discontinuously, as illustrated in the figure~\ref{fig:fopt}. The strength of the phase transition is characterized by the ratio $v_c/T_c$, where $v_c$ is the vacuum expectation value at the critical temperature, $T_c$. A value $v_c/T_c \gtrsim 1$ indicates a strong FOPT, whereas values approaching zero correspond to a SOPT. In the above estimation, the other couplings, such as $\lambda_{\eta,1,2}$, are assumed to be of order $\mathcal{O}(10^{-3})$ to ensure that their contributions are negligible.

\begin{figure*}[htbp!]
    \centering
    \includegraphics[width=0.85\linewidth]{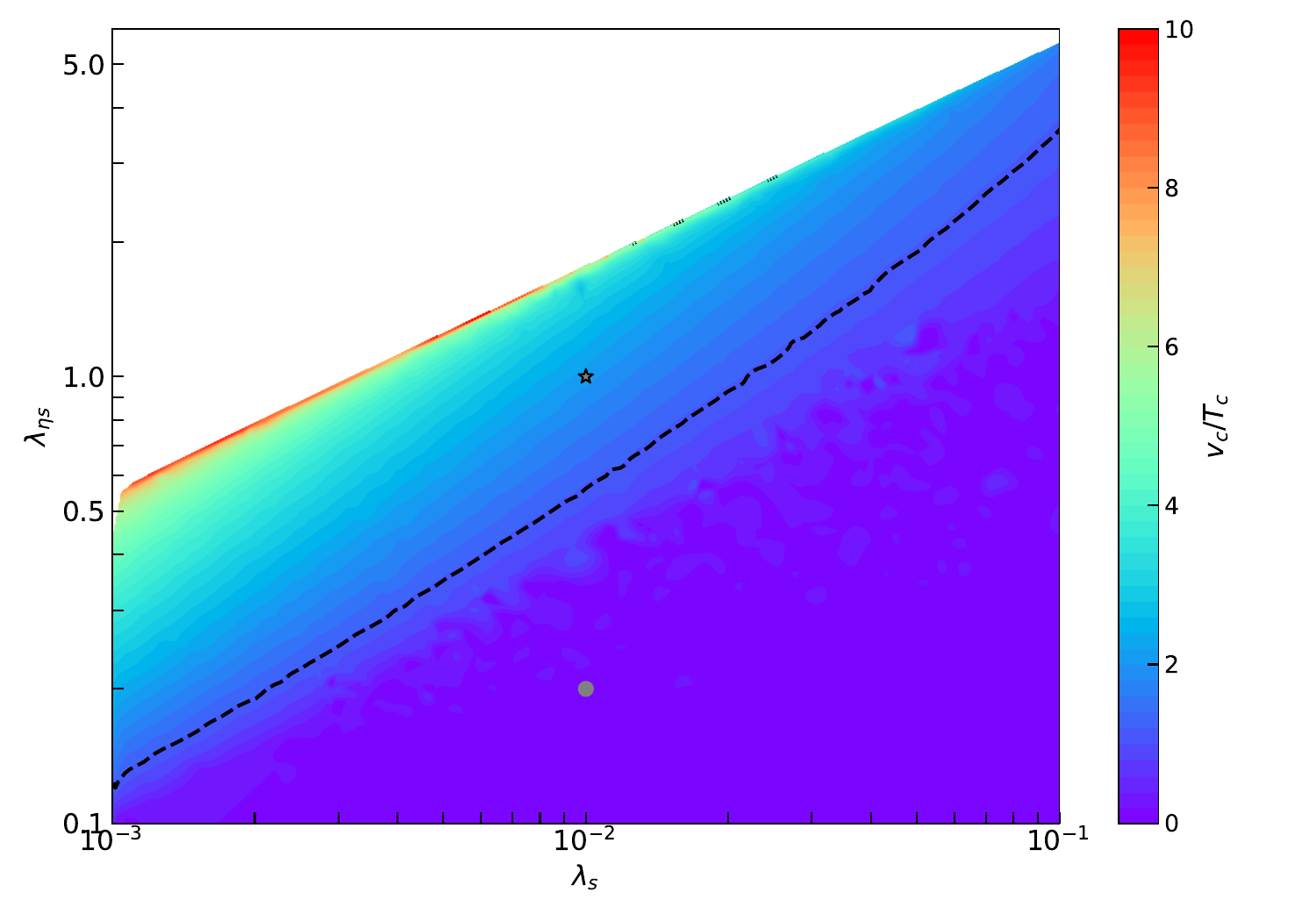}
    \caption{Parameter space in the $\lambda_s$--$\lambda_{\eta s}$ plane for $v_s = 10^6~\mathrm{GeV}$ and $\mu_\eta=10^{-2}v_s$. The dashed black line corresponds to the contour $v_c/T_c = 1$, indicating the boundary for a strong FOPT.}
    \label{fig:scan}
\end{figure*}

Before delving into the details of DW formation resulting from a FOPT, we would like to discuss the parameters that controls the strength of phase transition in the present setup. From figure~\ref{fig:sopt} and \ref{fig:fopt}, we find that the quartic interaction ($\lambda_{\eta s}$) between $\eta$ and $S$ modifies the thermal potential, thereby changing the phase transition from a SOPT to a FOPT scenario. To further examine the role of the self-quartic coupling of $S$ and its quartic interaction with $\eta$ in determining the nature of the phase transition, we show the corresponding parameter space in figure~\ref{fig:scan}. We observe that a significant region of the parameter space allows for FOPT for larger values of the quartic coupling $\lambda_{\eta s}$ up to a certain cutoff. Beyond this value, loop corrections modify the potential such that the symmetry remains unbroken at zero temperature. On the other hand, smaller values of $\lambda_{\eta s}$ correspond to SOPT. The $\bigstar$ and $\bigcirc$ markers denote points corresponding to FOPT and SOPT, respectively. The dashed black line represents the contour $v_c/T_c = 1$. We would like to emphasize that the choice of $v_s$ presented above is merely one representative example.

As the phase transition progresses, bubbles grow and eventually collide with one another. When two bubbles containing the same vacuum state ($+v_s$ and $+v_s$ or $-v_s$ and $-v_s$) collide, the field configuration merges smoothly and the interior becomes a single domain. While DWs are formed after bubbles of $+v_s$ and $-v_s$ vacua collide. When these bubbles collide, their walls bounce off each other and slow down due to surface tension, continuously losing energy. The wall eventually settles into a stable field configuration once sufficient energy has been dissipated. The initial separation between bubbles determines how many times they bounce before stabilizing.

\section{Domain wall biased by quantum gravity}
\label{QG}

The subsequent evolution of these residual DWs depends on the bias between the approximately degenerate vacua that can be calculated after replacing the fields $S$ and $H$ with their individual vevs in Eq. \eqref{eq:delV}. We get a bias contribution to the potential of the form

\begin{eqnarray}
         V_{\rm bias}  &=& \frac{1}{\Lambda_{\rm QG}} (  v_s^5 +   \frac{v_s^3 v_h^2}{2} +   \frac{v_s v_h^4}{4}+ v_s^3 \eta^2 +  v_s \eta^4 ). 
        \label{eq:vbias}
    \end{eqnarray}

\noindent The introduction of the bias terms in the potential allows DWs to annihilate that results in the stochastic GW background generation. As discussed, we consider a large hierarchy $v_s>>v_h$ in our analysis, as a result, the dominant contribution to the $V_{\rm bias}$ can be approximated as

\begin{eqnarray}
        V_{\rm bias}  &\simeq& \frac{ v_s^5}{\Lambda_{\rm QG}}   . 
        \label{eq:vbias2}
    \end{eqnarray}

Before examining GW production from DW annihilation in detail, we first address constraints on $V_{\rm bias}$ arising from DW decay into SM particles. DWs with very small energy bias can exist for long periods, since the annihilation timescale scales inversely with the bias ($t_{\rm ann}\propto V_{\rm bias}^{-1}$). Demanding that DWs collapse before dominating the Universe's energy density establishes a lower bound on $|V_{\rm bias}|$ \cite{Saikawa:2017hiv}.
\begin{equation}
V_{\rm bias}^{1/4}>2.18\times10^{-5} {\rm GeV}~ C_{\rm ann}^{1/4}\mathcal{A}^{1/2}\left(\frac{\sigma^{1/3}}{10^3\rm GeV}\right)^{3/2},
\label{vbias_LB}
\end{equation}
where $C_{\rm ann}\simeq 2$ is a dimensionless constant, and $\sigma=\sqrt{\frac{8\lambda_s}{9}}v_s^3$ is the surface tension of the DW. Satisfying Eq. \eqref{vbias_LB} ensures DWs annihilate before dominating the energy density. However, if DWs decay to SM particles, these decay products can destroy light elements synthesized during BBN. To avoid this, DW annihilation must complete by $t_{\rm ann}\leq 0.01$ s, which imposes a further constraint on $V_{\rm bias}$:
\begin{equation}
V_{\rm bias}^{1/4}>5.07\times10^{-4} {\rm GeV}~ C_{\rm ann}^{1/4}\mathcal{A}^{1/4}\left(\frac{\sigma^{1/3}}{10^3\rm GeV}\right)^{3/4}.
\label{vbias_LB2}
\end{equation}

\section{Gravitational Waves Signatures} 
\label{GW}
\subsection{Gravitational waves from second order phase transition: annihilating domain walls}

In this section, we discuss the case where the phase transition proceeds through second order and the GW is solely produced from the annihilation of the DWs.
The production of the GW from DW annihilation has been investigated in detail, for example, see refs. \cite{Vilenkin:1981zs, Gelmini:1988sf, Larsson:1996sp, Hiramatsu:2013qaa, Hiramatsu:2012sc, Saikawa:2017hiv,Roshan:2024qnv,Bhattacharya:2023kws}.  Assuming that the DW annihilate instantaneously ($t=t_{\text{ann}}$) during the radiation-dominated era, the peak frequency $f_{p}$ and peak energy density spectrum $\Omega_{p}h^2$ of GW at present can be expressed as \cite{Saikawa:2017hiv, Chen:2020wvu}
 \begin{widetext}
 \begin{align}\label{fpeak}
     f_{p}&\simeq 3.75\times10^{-9}~\text{Hz}\times C_{\rm ann}^{-1/2}\mathcal{A}^{-1/2}\bigg(\frac{10^3~\text{GeV}}{\mathcal{\sigma}^{1/3}}\bigg)^{3/2}\bigg(\frac{V_{\text{bias}}^{1/4}}{10^{-3}~\text{GeV}}\bigg)^{2}\,,\nonumber\\
     \Omega_{p}h^2&\simeq 5.3\times10^{-20}\times\tilde{\epsilon}_{\text{GW}}~C_{\rm ann}^{2}\mathcal{A}^{4}\bigg(\frac{\sigma^{1/3}}{10^3~\text{GeV}}\bigg)^{12}\bigg(\frac{10^{-3}~\text{GeV}}{V_{\text{bias}}^{1/4}}\bigg)^{8},
 \end{align}
 \end{widetext}
 where $\tilde{\epsilon}_{\text{GW}}\simeq0.7$~\cite{Hiramatsu:2013qaa} denotes the fraction of energy radiated into GW and can be regarded as a constant in the scaling regime. The GW spectrum from DWs characteristically exhibits a broken power-law form. The break frequency is set by the annihilation time, while the peak amplitude is determined by the DW energy density, as shown in Eq. \eqref{fpeak}. To describe this spectrum, we employ the broken power-law parametrization from Refs.~\cite{Caprini:2019egz, NANOGrav:2023hvm}:
\begin{eqnarray}
 \Omega_{\rm GW}h^2_{} = \Omega_p h^2 \frac{(a+b)^c}{\left(b x^{-a / c}+a x^{b / c}\right)^c} \ ,
\label{eq:spec-par}
\end{eqnarray}
where $x \equiv f/f_p$, and $a$, $b$ and $c$ are real and positive parameters. Here the low-frequency slope $a = 3$ can be fixed by causality, while numerical simulations suggest $b \simeq c \simeq 1$~\cite{Hiramatsu:2013qaa}.


Apart from the constraint on DW from equations \eqref{vbias_LB} and \eqref{vbias_LB2}, a large amplitude of GW from DW annihilation can also contribute to extra effective number of relativistic species, $\Delta N_{\rm eff}$. This gives an upper bound on the GW energy density. Considering the Planck 2018 results on $\Delta N_{\rm eff}$ at the time of BBN, the constraint on GW energy density is given as \cite{Domenech:2020ssp}
\begin{eqnarray}
    \Omega_{\rm GW , BBN} < \frac{7}{8} \left(\frac{4}{11}\right)^{4/3} \Delta N_{\rm eff} \sim 0.05.
\end{eqnarray}
Hence, the maximum energy density of GW today which is allowed from $\Delta N_{\rm eff}$ bound can be written as 
\begin{widetext}
\begin{eqnarray}
    \Omega_{\rm GW}^{\rm \Delta N_{\rm eff}}h^2 = 0.39 \left(\frac{g_{*}(T_{\rm BBN})}{106.75}\right)^{-1/3} \Omega_{r,0}h^2 \Omega_{\rm GW , BBN}  \simeq 1.75\times 10^{-6},
\end{eqnarray}
\end{widetext}
where $\Omega_{r,0}h^2\sim 4.18\times10^{-5}$ represents current radiation energy density fraction. Therefore, to be consistent with this bound, the peak energy density of GW from DW, $\Omega_{p}h^2$, should be less than $ \Omega_{\rm GW}^{\rm \Delta N_{\rm eff}}h^2$.

\begin{table}[hbt]
		\begin{center}
			\vskip 0.5 cm
			\begin{tabular}{|c|c|c|c|c|}
				\hline
				BPs	&	$v_s$ [GeV] & $\Lambda_{\rm QG}$ [GeV] &  $\lambda_s$ & $\lambda_{\eta s}$  \\
				\hline
                \hline
				1      &  $10^{6}$     &  $10^{32}$ & 0.01 & 1.19 \\                 
				\hline
				2    & $10^{10}$ &$10^{28}$ & $0.01$  & 1.16\\                 
				\hline
                3    & $10^{7}$ &$10^{27}$ & $0.1$  & 4.39\\                 
				\hline
			\end{tabular}
		\end{center}  
		\caption{Three characteristic benchmark values of $v_s$ and
$\Lambda_{\rm QG}$ are listed along with the values $\lambda_s$ and $\lambda_{\eta s}$. }
		\label{BP}
	\end{table}
    
\begin{figure*}[htbp!]
    \centering
    \includegraphics[width=0.85\linewidth]{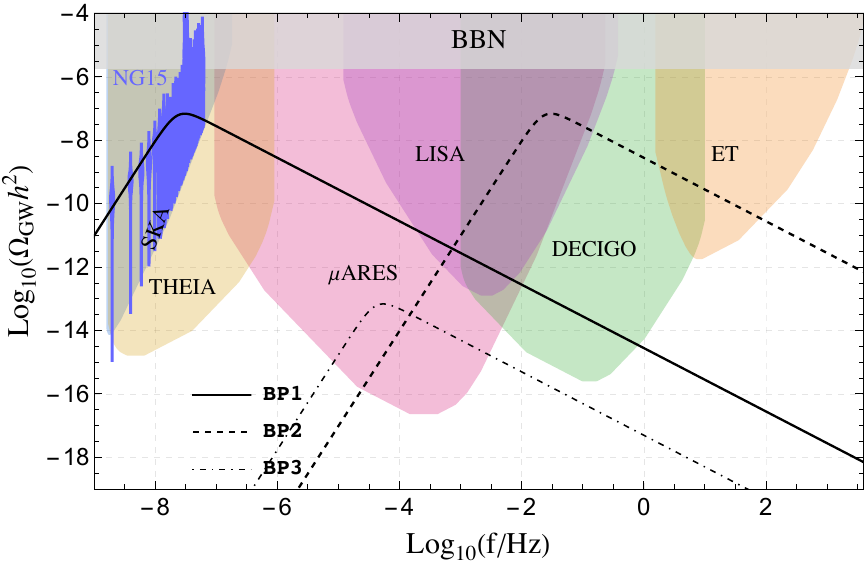}
    \caption{The gravitational wave spectrum produced from the annihilation of domain walls corresponds to the benchmark points listed in table~\ref{BP}.}
    \label{fig:DW_spec1}
\end{figure*}

In Fig. \ref{fig:DW_spec1}, we first present the power-law integrated sensitivity curves \cite{Thrane:2013oya} of the future GW detectors ET~\cite{Punturo:2010zz}, LISA~\cite{LISA:2017pwj}, DECIGO~\cite{Kawamura:2020pcg}, $\mu$Ares~\cite{Sesana:2019vho}, SKA~\cite{Janssen:2014dka}, and THEIA~\cite{Garcia-Bellido:2021zgu} which are evaluated following Eq.~\eqref{eq:spec-par} by calculating the signal-to-noise ratio (SNR)~\cite{Maggiore:1999vm,Allen:1997ad}
\begin{equation*}
    \varrho=\left[n_{\mathrm{det}} t_{\mathrm{obs}} \int_{f_{\min }}^{f_{\max }} d f\left(\frac{\Omega_{\text {signal }}(f)}{\Omega_{\text {noise }}(f)}\right)^2\right]^{1 / 2} \; ,
    \label{eq:SNR}
\end{equation*}
where $n_{\rm det} = 1$ for auto-correlated detectors and  $n_{\rm det} = 2$ for cross-correlated detectors, $t_{\rm obs}$ denotes the observational time, and $\Omega_{\text {noise }}$ represents the noise spectrum expressed in terms of the GW energy density spectrum~\cite{Schmitz:2020syl}. Integrating $(\Omega_{\rm signal}/\Omega_{\rm noise})^2$ over the sensitive frequency range of individual GW detectors, we obtain the SNRs for the GW spectra. Here we also choose
$\varrho = 10 $ as the threshold SNR. The gray shaded region (labeled BBN) represents the constraint $ \Omega_{\rm GW}^{\rm \Delta N_{\rm eff}}h^2$.

In Fig. \ref{fig:DW_spec1}, the curve corresponding to benchmark point (BP), BP1, we fix $v_s=10^6$ GeV, 
$\Lambda_{\rm QG}=10^{27}$ GeV and $\lambda_s=0.01$ such that the benchmark spectrum aligns with the NANOGrav 15-year result very well. Here the blue violins indicate the GW spectrum observed by the NANOGrav. Note that while $\lambda_{\eta s}$ does not affect the GW spectrum from DW annihilation, it plays a crucial role in the FOPT-generated GW analysis discussed in the next section. For BP2, we change $v_s=10^{10}$ GeV and 
$\Lambda_{\rm QG}=10^{28}$ GeV while keeping $\lambda_s$ fixed at 0.01. As expected, increasing the values of both $v_s$ and $\Lambda_{\rm QG}$ increases $\Omega_{p}h^2\propto (\lambda_s^2 \Lambda_{\rm QG}^2)/v_s^2$ as well as $f_p\propto v_s/(\lambda_s^{1/4}\Lambda_{\rm QG}^{1/2})$, this is visible from the black dashed line in Fig. \ref{fig:DW_spec1}. Finally, we have chosen BP3 in such a way that the DM phenomenology discussed in section \ref{DMPH} can also be probed by the GW experiments. These BPs also obey the constraints from Eqs. \eqref{vbias_LB} and \eqref{vbias_LB2}.

\subsection{Gravitational waves from first order phase transition: a twin-peaked spectrum}

As detailed in Section \ref{DW_FOPT}, DWs form during a FOPT when bubbles of $+v_s$ and $-v_s$ collide. In our scenario, QG effects introduce a bias, so DWs' late-time evolution occurs in a thermal plasma consisting entirely of the true vacuum. Importantly, the bias in the scalar potential must be chosen such that the FOPT completes on a timescale much shorter than the Hubble time.

\begin{figure*}[htbp!]
    \centering
    \includegraphics[width=0.85\linewidth]{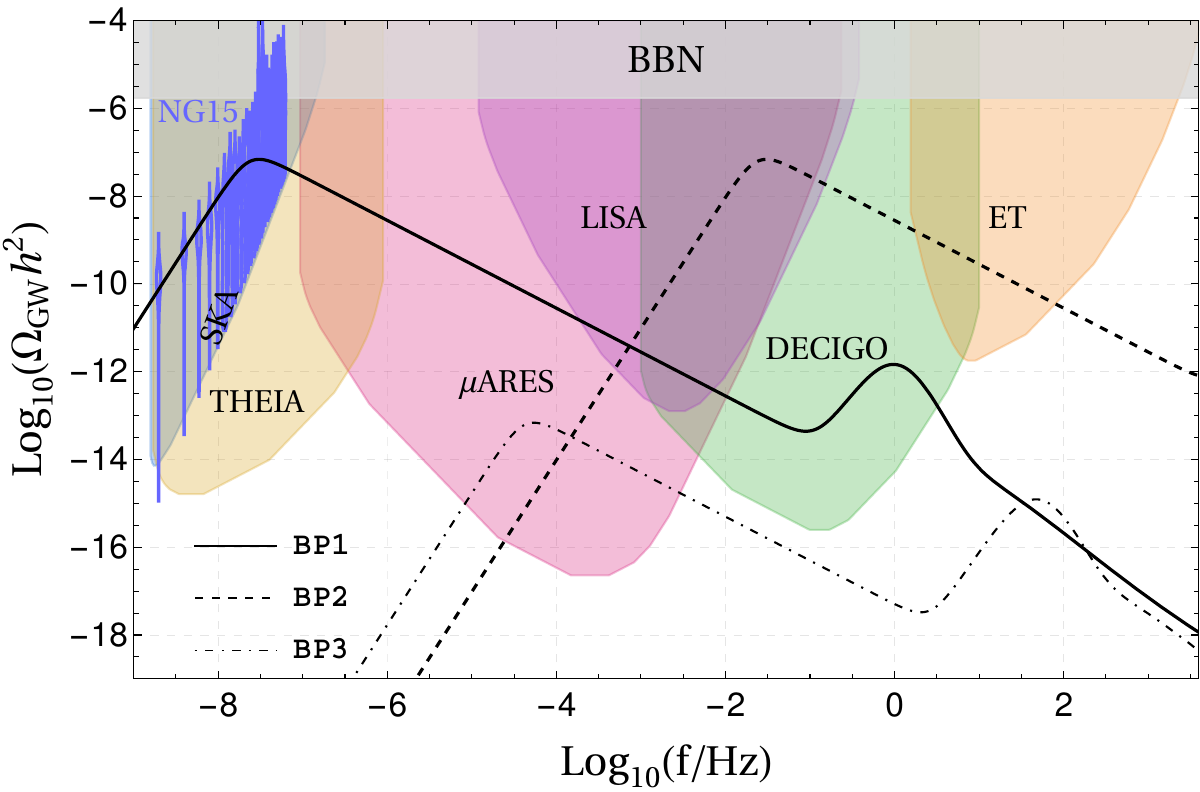}
    \caption{The combined gravitational wave spectrum arising from the FOPT and the annihilation of domain walls corresponds to the benchmark points listed in table~\ref{BP} and \ref{BP_gw}.}
    \label{fig:DW_PT_spec1}
\end{figure*}

While GW production from annihilating DWs proceeds as discussed in the previous section, an additional GW source emerges here: GW from the FOPT itself. This produces an interesting twin-peaked GW spectrum (see Fig. \ref{fig:DW_PT_spec1}). We now discuss the GW production during the FOPT resulting from the bubble collisions of same type vacua~\cite{Turner:1990rc,Kosowsky:1991ua,Kosowsky:1992rz,Kosowsky:1992vn,Turner:1992tz}, the sound wave of the plasma~\cite{Hindmarsh:2013xza,Giblin:2014qia,Hindmarsh:2015qta,Hindmarsh:2017gnf} and the turbulence of the plasma~\cite{Kamionkowski:1993fg,Kosowsky:2001xp,Caprini:2006jb,Gogoberidze:2007an,Caprini:2009yp,Niksa:2018ofa}. One should note that in this scenario, an additional contribution to the GW spectrum from the FOPT arises from collisions between bubbles of the opposite vacuum which do not merge (i.e. $+v_s$ and $-v_s$  ). However, as shown in \cite{Wei:2022poh}, once these bubble collisions complete, their GW contribution becomes subdominant and even negligible compared to the DW network, which becomes the primary GW source during later evolution. Hence we neglect this contribution to the GW arising from collisions between bubbles of the opposite vacuum throughout our analysis. The total GW spectrum is then given by 
$$\Omega_{\rm GW}^{\rm PT}(f) = \Omega_s(f) + \Omega_{\rm sw}(f) + \Omega_{\rm turb}(f).$$ 
While the peak frequency and peak amplitude of such GW spectrum depend upon specific FOPT related parameters, the exact nature of the spectrum is determined by numerical simulations. Before going into the details of the GW spectrum from all three sources mentioned above, we first discuss the key parameters relevant for GW estimates namely, the inverse duration of the phase transition and the latent heat released are calculated and parametrised in terms of \cite{Caprini:2015zlo}
$$\frac{\beta}{{ H}(T)} \simeq T\frac{d}{dT} \left(\frac{S_3}{T} \right) $$ and 
$$ \alpha_* =\frac{1}{\rho_{\rm rad}}\left[\Delta V_{\rm tot} - \frac{T}{4} \frac{\partial \Delta V_{\rm tot}}{\partial T}\right]_{T=T_n} $$ 
respectively, where $S_3$  is the bounce action for an $ O(3)$ symmetric bounce solution and $\Delta V_{\rm tot}$ is the energy difference in true and false vacua. The bubble wall velocity $v_w$ is estimated from the Jouguet velocity \cite{Kamionkowski:1993fg, Steinhardt:1981ct, Espinosa:2010hh}
$$v_J = \frac{1/\sqrt{3} + \sqrt{\alpha^2_* + 2\alpha_*/3}}{1+\alpha_*}$$
following the prescription given in \cite{Lewicki:2021pgr}. 
 The GW spectrum for bubble collision is given by \cite{Caprini:2015zlo}
\begin{widetext}
\begin{equation}
    \Omega_s h^2 = 1.67 \times 10^{-5} \left ( \frac{100}{g_*} \right)^{1/3} \left(\frac{{ H_*}}{\beta}\right)^2 \left(\frac{\kappa_s \alpha_*}{1+\alpha_*}\right)^2 \frac{0.11 v^3_w}{0.42+v^2_w} \frac{3.8(f/f_{\rm peak}^{\rm PT, s})^{2.8}}{1+2.8 (f/f_{\rm peak}^{\rm PT, s})^{3.8}}\,,
\end{equation}
\end{widetext}
where the peak frequency \cite{Caprini:2015zlo} is 
\begin{eqnarray}   
f_{\rm peak}^{\rm PT, s} &=& 1.65 \times 10^{-5} {\rm Hz} \left ( \frac{g_*}{100} \right)^{1/6} \left ( \frac{T_n}{100 \; {\rm GeV}} \right ) \nonumber \\
   &\times&  \frac{0.62}{1.8-0.1v_w+v^2_w} \left(\frac{\beta}{{ H_*}}\right).
\end{eqnarray}
The Hubble expansion parameter at nucleation temperature is denoted as $H_*=H(T_n)$. The efficiency factor $\kappa_s$ for bubble collision can be expressed as \cite{Kamionkowski:1993fg}
\begin{align}
    \kappa_s=\frac{1}{1+0.715 \alpha_*}\left(0.715\alpha_* +\frac{4}{27}\sqrt{3\alpha_*/2}\right).
\end{align}
Subsequently, the collision of bubble walls with the surrounding plasma leads to the generation of sound waves. The GW spectrum produced from the sound wave in the plasma can be written as \cite{Caprini:2015zlo,Caprini:2019egz,Guo:2020grp}
\begin{widetext}
\begin{equation}
    \Omega_{\rm sw} h^2 = 2.65 \times 10^{-6} \left ( \frac{100}{g_*} \right)^{1/3} \left(\frac{{ H_*}}{\beta}\right) \left(\frac{\kappa_{\rm sw} \alpha_*}{1+\alpha_*}\right)^2 v_w (f/f_{\rm peak}^{\rm PT, sw})^{3} \left(\frac{7}{4+3 (f/f_{\rm peak}^{\rm PT, sw})^{2}} \right)^{7/2} \Upsilon
\end{equation}
\end{widetext}
and the corresponding peak frequency is given by \cite{Caprini:2015zlo}
\begin{eqnarray}
    f_{\rm peak}^{\rm PT, sw} &=& 1.65 \times 10^{-5} {\rm Hz} \left ( \frac{g_*}{100}\right)^{1/6}\frac{1}{v_w}  \left ( \frac{T_n}{100 \; {\rm GeV}} \right )\nonumber \\  &\times&\left(\frac{\beta}{{ H_*}}\right) \frac{2}{\sqrt{3}}.
\end{eqnarray}
The efficiency factor for sound wave can be expressed as \cite{Espinosa:2010hh}
\begin{align}
    \kappa_{\rm sw}=\frac{\sqrt{\alpha_*}}{0.135+ \sqrt{0.98+\alpha_*}}.
\end{align}
The suppression factor is given by
$\Upsilon = 1-\frac{1}{\sqrt{1+2\tau_{\rm sw}H_*}},$
which depends on the lifetime of the sound waves, $\tau_{\rm sw}$~\cite{Guo:2020grp}. The sound-wave lifetime can be estimated as $\tau_{\rm sw} \sim R_*/\bar{U}_f$, where the mean bubble separation is $R_*=(8\pi)^{1/3} v_w \beta^{-1}$ and the root-mean-square fluid velocity is $\bar{U}_f=\sqrt{3\kappa_{\rm sw}\alpha_*/4}$. Finally, the gravitational wave spectrum generated by magnetohydrodynamic turbulence in the plasma is given by~\cite{Caprini:2015zlo}.
\begin{widetext}
\begin{equation*}
    \Omega_{\rm turb} h^2 = 3.35 \times 10^{-4} \left ( \frac{100}{g_*} \right)^{1/3} \left(\frac{{ H_*}}{\beta}\right) \left(\frac{\kappa_{\rm turb} \alpha_*}{1+\alpha_*}\right)^{3/2} v_w \frac{(f/f_{\rm peak}^{\rm PT, turb})^{3}}{(1+ f/f_{\rm peak}^{\rm PT, turb})^{11/3} (1+8\pi f/h_*)}
\end{equation*}
\end{widetext}
with the peak frequency as \cite{Caprini:2015zlo}
\begin{eqnarray}
    f_{\rm peak}^{\rm PT, turb} &=& 1.65 \times 10^{-5} {\rm Hz} \left ( \frac{g_*}{100} \right)^{1/6}\frac{1}{v_w}  \left ( \frac{T_n}{100 \; {\rm GeV}} \right )\nonumber\\ &\times&\frac{3.5}{2} \left(\frac{\beta}{{H_*}}\right).
\end{eqnarray}
The efficiency factor for turbulence is taken to be $\kappa_{\rm turb} \simeq 0.1\,\kappa_{\rm sw}$. 
The inverse Hubble time at the epoch of gravitational wave production, redshifted to the present day, is given by
\begin{equation}
   h_*= 1.65\times 10^{-5} \frac{T_n}{100 \hspace{0.1 cm} \rm GeV} \left(\frac{g_*}{100}\right)^{1/6}.
\end{equation}
 It is clear from the above expressions that the contribution from sound waves turns out to be the most dominant one and the peak of the total GW spectrum corresponds to the peak frequency of sound waves contribution.

 \begin{table}[hbt]
		\begin{center}
			\vskip 0.5 cm
			\begin{tabular}{|c|c|c|c|c|c|}
				\hline
				BPs	&	$T_c$ [GeV] & $v_c$ [GeV] &  $T_n$ [GeV]&  $\alpha_*$ & $\beta/H_*$ \\
				\hline
                \hline
				1      &  $3.7\times10^{5}$     &  $9.2\times10^{5}$ & $1.5\times10^{5}$  & 0.06 & 69.6 \\                 
				\hline
				2    &  $3.8\times10^{9}$     &  $9.1\times10^{9}$ & $1.6\times10^{9}$  & 0.06 & 97.2 \\                 
				\hline
                3    &  $8.1\times10^{6}$     &  $9.9\times10^{6}$ & $4.9\times10^{6}$  & 0.01 & 107.1 \\               
				\hline
			\end{tabular}
		\end{center}  
		\caption{FOPT parameters and other details involved in the GW spectrum calculation correspond to Table~\ref{BP}. }
		\label{BP_gw}
	\end{table}


 The contribution to the GW spectrum coming from the FOPT can be seen in Fig. \ref{fig:DW_PT_spec1}. The twin peaks observed in Fig. \ref{fig:DW_PT_spec1} have distinct physical origins: the higher-frequency peak traces GW production during the FOPT, while the lower-frequency peak comes from DW annihilation. Since DWs persist and annihilate only after the phase transition completes, their GW emission occurs later and thus appears at lower frequency today.
 
 This scenario also has a compelling observational feature as visible from the spectrum corresponding to BP1 in Fig. \ref{fig:DW_PT_spec1}: a single symmetry breaking generates GW detectable by both PTAs and interferometers. A detection by one experiment would immediately point to complementary signals in the other. In addition to this, we also show two other BPs in the figure. While BP2 exhibits only the DW peak in the observable range (with the FOPT peak at higher frequency), BP3 is specifically tuned to satisfy DM phenomenology: the candidate $\chi$
reproduces the measured relic density and remains accessible to complementary DM detection experiments. We provide an elaborate discussion of the DM phenomenology in Section \ref{DMPH}. 

\section{Dark matter Phenomenology}
\label{DMPH}
\subsection{Freeze-in Dark matter}

The present set-up provides an option of producing the DM via freeze-in mechanism. Once the $\ZDW$ is spontaneously broken, while $\eta$ obtains a  mass  $m_{\eta}^2=\mu_\eta^2+\lambda _{\eta s}v_s^2$, the DM mass remains unaffected, $m_\chi$. If kinematically allowed $m_\eta>  m_\chi$, $\eta$ can decay and produce $\chi$ and the SM leptons. Assuming the DM production happen through the freeze-in mechanism, the coupling $y_\chi$ has to be extremely feeble in order to prohibit DM from thermalizing with the surrounding plasma. In such a case, the DM abundance can be estimated using~\cite{Barman:2020plp}
\begin{equation}
   Y_{\text{DM}}\simeq \frac{3\times 135\, M_{\text{Pl}}}{1.66\times 8\pi^3g^S_\star\sqrt{g_\star^\rho}}\frac{\Gamma_{\eta\to\chi\ell}}{m_\eta^2} \; , 
   \label{DM_abundance}
\end{equation}
\noindent where the decay rate assuming $m_\eta>>m_\chi$ is given as, 
\begin{equation}
    \Gamma_\eta=\frac{y_\chi^2}{16\pi}m_\eta,
\end{equation}
\noindent and the factor 3 in numerator of Eq. \ref{DM_abundance} comes from the fact that all the component of $\eta$ decay to the DM with same rate. Following this, one can obtain the DM relic density at the present epoch using
\begin{equation}
    \Omega_{\text{DM}} h^2=2.75\times 10^8 \bigg(\frac{m_{\text{DM}}}{\text{GeV}}\bigg) Y_{\text{DM}} \; .
\end{equation}
The region of parameter space producing the observed DM relic density $\Omega_{\rm DM}h^2=0.12\pm 0.0012$ \cite{Planck:2018vyg} for different values of $y_\chi$ is shown by the red and blue solid lines in Fig. \ref{DMplt} as an illustrative example. We also show the gray-colored region excluded from the Lyman-$\alpha$ bound  on freeze-in DM \cite{Decant:2021mhj}. 
\subsection{Dark matter decay via quantum gravity effects}
\label{sec3}

After the breaking of EW symmetry,  eq. \eqref{eq:Z2both} can be expressed in terms of the vevs $v_s$ and $v_h$  as 
\begin{eqnarray}
        \mathcal{L}_{\text{break}}  &=& \sum_{i=1,2,3} \frac{ \beta_iv_s v_h}{\sqrt{2}\Lambda_{\rm QG}}  \overline{\nu_{L_i}}\chi  \;.
        \label{eq:Z2both2}
    \end{eqnarray}
    This generates a Dirac mass term $m_{D_i}= \frac{ \beta_i v_s v_h}{\sqrt{2}\Lambda_{\rm QG}}$ for the SM neutrinos identical to the one obtained in the Type-I seesaw mechanism \cite{Minkowski:1977sc,Yanagida:1979as,Yanagida:1979gs,GellMann:1980vs,Mohapatra:1979ia,Schechter:1980gr,Schechter:1981cv}. 
Consequently, the SM neutrino mixes with the DM $\chi$. The mixing angle $\theta$ in this situation can then be written as \cite{Datta:2021elq}, 
\begin{equation}
\theta
 \simeq\sum_{i=1,2,3}\left(\frac{m_{D_i}}{m_{\text{DM}}}\right)
 =\sum_{i=1,2,3}\left(\frac{\beta_i v_sv_h}{\sqrt{2}\Lambda_{\text{QG}}m_{\text{DM}}} \right)\; .   
 \label{eq:mixingangle}
\end{equation}

\noindent The assumption  $\beta_1=\beta_2=\beta_3=1$ discussed earlier also suggest that the DM interacts with different generations of neutrinos with the identical ${\cal O}(1)$  strength hence one can write,
\begin{equation}
\theta
 \simeq\frac{3 v_sv_h}{\sqrt{2}\Lambda_{\text{QG}}m_{\chi}} \; .   
 \label{eq:mixingangle_new}
\end{equation}

This mixing allows the DM to decay to the SM particles and provides us a possibility to look for DM in the indirect search experiments. For example, the mixing allows the DM to have two-body (one-loop level mediated via $W^\pm$ and SM charged leptons) and three-body (at tree-level mediated via $W^\pm$) decay channels that can dominate if kinematically allowed. At one-loop, $\chi$ can decay to photons and active neutrinos via $\chi \to \nu \gamma$ with the decay rate~\cite{Shrock:1982sc,Essig:2013goa}
\begin{equation}
\begin{split}
\tau_{\chi \rightarrow \nu \gamma} & \simeq\left(\frac{9 \alpha_{\mathrm{EM}} \sin ^2 \theta}{1024 \pi^4} G_F^2 m_\chi^5\right)^{-1} \\
& \simeq 1.8 \times 10^{17}~{\rm s}\left(\frac{10 \, \mathrm{MeV}}{m_\chi}\right)^5\left(\frac{\sin \theta}{10^{-8}}\right)^{-2} \; ,
\end{split}
\label{eq:lifetime}
\end{equation}
where $\alpha_{\rm EM} = 1/137$ is the electromagnetic fine-structure constant and $G_F$ denoting the Fermi constant. The decay rate  of three-body decay channel $\chi \to e^+ e^- \nu$ can be approximately expressed as \cite{Ruchayskiy:2011aa,Essig:2013goa}
\begin{equation}
\begin{split}
    \tau_{\chi \to e^+ e^- \nu} & \simeq \left(   \frac{c_\alpha \sin^2 \theta}{96 \pi^3} G^2_F m^5_{\chi}\right)^{-1} \\
    & \simeq 2.4 \times 10^{15}~ {\rm s} \left(\frac{10 \mathrm{MeV}}{m_\chi}\right)^5\left(\frac{\sin \theta}{10^{-8}}\right)^{-2} \;, 
    \end{split}
    \label{eq:lifetime3}
\end{equation}
where $c_{\alpha}=(1+ 4 \sin^2 \theta_W + 8 \sin^4 \theta_W)/4 \simeq 0.59$ with $\theta_W$ being the weak mixing angle.
Contributions from the above two channels to the X/$\gamma$-ray fluxes are roughly at similar levels~\cite{Essig:2013goa}. The null detection of X-rays and $\gamma$-rays line set a lower bound on the lifetime of the decaying DM that can further be converted into the constraints on the DM mass and the mixing angle. The observed diffuse photon spectra data obtained from the HEAO-1~\cite{Gruber:1999yr}, INTEGRAL~\cite{Bouchet:2008rp}, COMPTEL~\cite{Sreekumar:1997yg} and EGRET~\cite{Strong:2003ey} satellites can restrict the parameter space of $m_\chi$ and $\theta$ within the range $0.01~{\rm MeV} \lesssim m_{\chi} \lesssim 100~{\rm MeV}$, which can be parameterized as $\theta^2 \lesssim 2.8 \times 10^{-18}({\rm MeV}/m_{\chi})^5$~\cite{Boyarsky:2009ix}. With  the help of the expression of $\theta$ in Eq.~\eqref{eq:mixingangle}, we have
\begin{equation}
\left(\frac{3 v_s v_h}{\sqrt{2}\Lambda_{\text{QG}}m_\chi}\right)^2 \lesssim 2.8\times10^{-18}\left(\frac{\text{MeV}}{m_\chi}\right)^5 \; ,
\label{DM-decay}
\end{equation}
where we consider that the DM interacts with different generations of neutrinos with the identical ${\cal O}(1)$  strength. Additionally, the Fermi Large Area Telescope (Fermi-LAT)~\cite{Fermi-LAT:2012pls,Fermi-LAT:2012ugx} has presented a dedicated line search of the diffuse $\gamma$-ray background, the null result of which also constrains the lifetime of the DM within the mass range $1~{\rm GeV} \lesssim m_\chi \lesssim 1~{\rm TeV}$.  Here we take the $95\%$ C.L. limit on the lifetime of decaying DM given by ref.~\cite{Fermi-LAT:2015kyq}, which assumes a Navarro-Frenk-White (NFW) profile for the DM distribution~\cite{Navarro:1995iw}. 

\begin{figure*}[htbp!]
    \centering
    \includegraphics[width=0.55\linewidth]{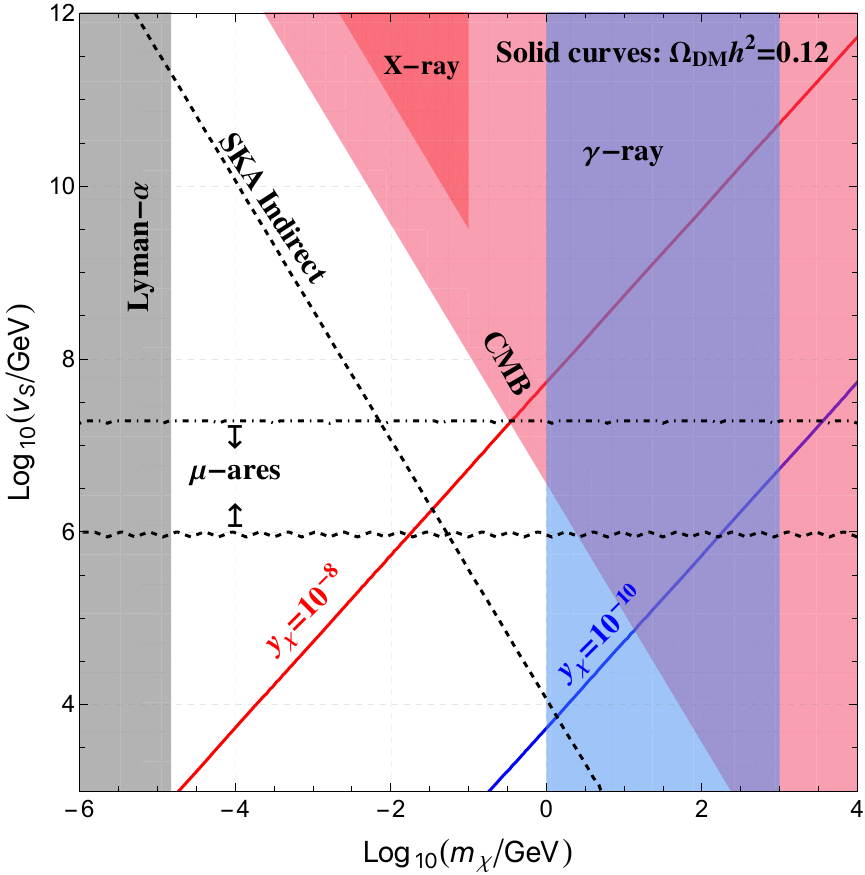}
    \caption{Constraints on the DM mass $m_\chi$ and the VEV $v_s$ of the scalar field from the DM relic density and indirect detections for $\Lambda_{\rm QG}=10^{27}$ GeV, $\mu_\eta=10^{-2}v_s,~\lambda_{\eta s}=4.39$. The blue, and red solid curves denote the
correlations between $m_\chi$ and $v_s$ we should have to satisfy the correct DM relic density $\Omega_{\rm DM}h^2 = 0.12$ for different $y_\chi=10^{-8},10^{-10}$. The blue, red, pink,
and gray-shaded areas represent respectively the excluded regions by the $\gamma-$ray decay, X-ray, CMB, and Lyman-$\alpha$
observations.}
    \label{DMplt}
\end{figure*}

A decaying DM also faces stringent constraints from CMB, if the decay occurs after or during the recombination. Following the decay, the decay products can re-ionize the intergalactic medium, and can modify the CMB power spectrum.  Accurate measurements of the CMB spectrum have been implemented by recent experiments including WMAP~\cite{WMAP:2012nax}, ACT~\cite{ACTPol:2014pbf}, SPT~\cite{Hou:2012xq} and Planck~\cite{Planck:2018vyg}. The $95\%$ C.L. lower bounds on the DM decay lifetime have been given in ref.~\cite{Slatyer:2016qyl}, showing that $\tau_{\rm DM} \gtrsim 10^{25}~{\rm s}$ for decays into both $e^+e^-$ pairs and photons. These DM decays can also introduce radio signals originating inside the DM-dominated galaxies and clusters, if the produced $e^+e^-$ pairs undergo energy loss via electromagnetic interactions in the interstellar medium. Such radio waves can be tested by radio telescopes like the Square Kilometer Array (SKA) radio telescope~\cite{Colafrancesco:2015ola}. It is found that the DM decay width up to $\Gamma_\text{DM}\gtrsim10^{-30}~{\rm s}^{-1}$ is detectable at SKA assuming  100-hour observation time~\cite{Dutta:2022wuc}. We incorporate all these constraints in Fig. \ref{DMplt} while showing a region of allowed parameter space for a fixed value of $\Lambda_{\rm QG}=10^{27}$ GeV, $\mu_\eta=10^{-2}v_s,~\lambda_{\eta s}=4.39$. Note that the choice of parameters presented above is merely one illustrative example.

\section{Conclusion}
\label{conclusion}
This work continues our previous studies \cite{King:2023ayw,King:2023ztb,Borah:2024kfn}, with a new ingredient, an additional $SU(2)_L$ scalar doublet. This scalar helps enable a phase transition while simultaneously producing fermionic DM through its decay via the freeze-in mechanism. Both scalars ($S$ and $\eta$) are odd under the discrete global $\ZDW$ symmetry, while $\eta$ additionally carries a nontrivial charge under $\ZDM$, which stabilizes the DM $\chi$ (also odd under $\ZDM$) on cosmological timescales. The scalar $S$ is responsible for breaking the $\ZDW$ symmetry and creating the DW. However, both $\Ztwo$
symmetries are explicitly broken by non-perturbative QG effects, leading to observable signatures from DM decay and GW from DW annihilation. While SOPT generates a single-peak GW spectrum resulting from DW annihilation, FOPT produces a twin-peak spectrum: one from DW annihilation and an additional higher-frequency peak arising directly from the phase transition dynamics.

This study has a compelling multi-messenger feature: a single symmetry breaking generates GW detectable by both pulsar timing arrays and interferometers, with detection by one experiment pointing to complementary signals in the other. This makes collaboration between GW experimental communities essential to map the complete spectrum. Furthermore, the framework naturally connects GW observations with DM searches, as the same physics drives both phenomena. Joint constraints from GW detectors and DM experiments significantly enhance predictive power and help resolve degeneracies in identifying early universe sources, a persistent challenge when relying on GW data alone. While GW experiments can confirm the occurrence of phase transitions and DW networks, complementary constraints from DM experiments can pin down the specific parameter space and discriminate between competing cosmological scenarios that might produce similar GW spectra.

\section{Acknowledgment}			
Authors would like to thank Graham White, Rinku Maji and Qaisar Shafi for useful discussions. RR acknowledges financial support from the STFC Consolidated Grant ST/T001011/1. The work of I.S. is supported by SERB, Government of India grant CRG/2022/000603.
\bibliographystyle{apsrev4-1}
\bibliography{Journal_v2.bbl}

\onecolumngrid
\appendix

\end{document}